\documentclass[aps,letterpaper,superscriptaddress,twocolumn,longbibliography,nofootinbib]{revtex4-1}
\pdfoutput=1

\usepackage{bmpsize}

\usepackage{hyperref}
\usepackage{color}
\usepackage{graphicx}	
\usepackage[utf8]{inputenc} 
\usepackage{bm}
\usepackage{amsmath}
\usepackage{amsfonts}

\usepackage{eufrak}
\def\sumint{\,\hbox{$\sum$}\!\!\!\!\!\!\!\int}

\begin{document}

\title{Analytic Transport from Weak to Strong Coupling in the O(N) model}

\author{Paul Romatschke}
\affiliation{Department of Physics, University of Colorado, Boulder, Colorado 80309, USA}
\affiliation{Center for Theory of Quantum Matter, University of Colorado, Boulder, Colorado 80309, USA}

\begin{abstract}
   In this work, a second-order transport coefficient (the curvature-matter coupling $\kappa$) is calculated exactly for the O(N) model at large N for any coupling value. Since the theory is ``trivial'' in the sense of possessing a Landau pole, the result for $\kappa$ only is free from cut-off artifacts much below the Landau pole in the effective field theory sense. Nevertheless, this leaves a large range of coupling values  where this transport coefficient can be determined non-perturbatively and analytically with little ambiguity. Along with thermodyamic results also calculated in this work, I expect exact large N results to provide good quantitative predictions for N=1 scalar field theory with $\phi^4$ interaction.
\end{abstract}

\maketitle
\section{Introduction}
Transport coefficients determine the real-time relaxation of a perturbation around a state of equilibrium. Familiar transport coefficients include conductivities, diffusion coefficients and viscosities. However, these well-known transport coefficients merely approximate the response of a system to a perturbation through a linear (first order) relationship with the local gradient. In real systems, there are non-linear corrections (second order, third order, etc.) which come with their own respective transport coefficients. For many applications, ignoring these higher-order terms constitutes a reasonable approximation, but for some perturbations, in particular those where gradients are strong, knowledge of second-order transport coefficients is important. Also, there are different types of perturbations (``channels'') which predominately couple to different combinations of transport coefficients, for instance the sound channel (longitudinal compression mode coupling to shear and bulk viscosity) and the shear channel (coupling predominantly to shear viscosity). Sometimes relations of transport coefficients between different channels exist, such as the well-known Einstein relation between the diffusion coefficient and conductivity.

For the purpose of this work, I will consider the somewhat exotic transport coefficient $\kappa$, which appears as second-order correction in the familiar sound and shear mode channels, and which was introduced in Refs.~\cite{Bhattacharyya:2008jc,Baier:2007ix} in the context of relativistic fluid dynamics. However, $\kappa$ enters into the description of relativistic fluid as the \textit{leading} order correction when considering the coupling of matter to perturbations in the curvature of space-time (e.g. gravitational waves). In the hydrodynamic gradient expansion of the energy momentum tensor
\begin{equation}
  \label{eq:grad-exp}
  T^{\mu\nu}=T^{\mu\nu}_0+T^{\mu\nu}_1+T^{\mu\nu}_2+\ldots\,,
\end{equation}
this comes about because the second-order term $T_2^{\mu\nu}$ includes contributions such as \cite{Romatschke:2017ejr} 
\begin{equation}
  \label{eq:t2}
  T_2^{\mu\nu}=\kappa R^{<\mu\nu>}+\ldots\,,
\end{equation}
where $R^{\mu\nu}$ is the Ricci tensor and $\langle\rangle$ denotes symmetric traceless projection. Since the Ricci tensor is second-order in a gradient expansion, this shows that $\kappa$ is the leading order transport coefficient for gravity-matter perturbations.
  
Because of relations similar in nature to the Einstein relations for diffusion, this curvature-matter coupling coefficient $\kappa$ also enters in the real-time evolution of sound waves in \textit{flat} space-time (albeit as a correction to first-order transport governed by shear and bulk viscosity). Therefore, even though $\kappa$ predominantly governs the interactions between space-time curvature and matter, this transport coefficient can be calculated by considering correlation functions in flat space-time (``Kubo formulas''). Results for $\kappa$ are currently available for free field theory \cite{Romatschke:2009ng,Moore:2012tc}, infinitely strongly coupled gauge theories in the limit of large 't Hooft coupling and larger number of colors \cite{Bhattacharyya:2008jc,Baier:2007ix,Finazzo:2014cna,Grozdanov:2016fkt}, and SU(3) gauge theory from lattice simulations \cite{Philipsen:2013nea}.

With the exception of the numerical constraints from Ref.~\cite{Philipsen:2013nea}, $\kappa$ is unknown in any quantum field theory except near coupling values of $\lambda\simeq 0,\infty$. Given that $\kappa$ is hardly of crucial relevance in most transport applications, one might be tempted to blame this apparent lack of knowledge on an apparent lack of interest. 

Unfortunately, the situation is hardly better for other, more familiar transport coefficients which \textit{are} of crucial importance in most transport situations. For instance, for scalar field theory and QCD the shear viscosity coefficient has been calculated in perturbation theory around vanishing coupling in Refs.~\cite{Jeon:1994if,Arnold:2003zc,Moore:2007ib,Ghiglieri:2018dib}, and in large N gauge theories near infinite coupling in Refs.~\cite{Policastro:2001yc,Kats:2007mq,Buchel:2008sh}. At intermediate coupling, results exist for QED in the limit of a large number of fermions \cite{Moore:2001fga,Aarts:2005vc} and for SU(3) gauge theory there are constraints from lattice simulations  \cite{Meyer:2007ic,Pasztor:2018yae}.

So why focus on calculating exotic transport coefficients when there is such need for the shear viscosity? The answer is that $\kappa$ is considerably easier to calculate because it can be extracted from \textit{Euclidean} (imaginary-time) rather than \textit{retarded} (real-time) correlation functions. However, there may be hope to generalize the calculation presented here to other transport coefficients.

In this work, I calculate $\kappa$ for a particular theory (the O(N) model with quartic interactions) where such a transport calculation is feasible. Somewhat unfortunately, in the large N limit the O(N) model in 3+1 dimensions possesses a positive $\beta$-function for all coupling values. Integrating the $\beta$ function, the coupling diverges at a finite energy scale (aka the Landau pole).  The theory is thus UV-incomplete or ``trivial''.  For energy scales close to the Landau pole, all possible irrelevant operators contribute, and hence observables will be sensitive to the particular discretization (the form of the Lagrangian) chosen for the theory. However, a (non-perturbative) renormalization program can be carried through for IR-safe observables such as the pressure, and UV-incomplete theories may be interpreted as effective low-energy descriptions. Thus, the O(N) model may be considered phenomenologically viable at energy scales well below the Landau pole. In practice, sensitivity to the cutoff scale can be tested for by varying the renormalization scale parameter, thus providing a quantitative handle on the breakdown of the theory.

\section{The Calculation}

Hydrodynamics provides the universal low energy/long wavelength description of matter. As such, hydrodynamics can be set up from a gradient expansion and the symmetries of the system under consideration, and universally determines the form of the n-point functions of the energy-momentum tensor $T^{\mu\nu}$, cf. Ref.~\cite{Romatschke:2017ejr}. Using a construction valid up to (including) second order in gradients (\ref{eq:grad-exp}), variation of the full $T^{\mu\nu}$ with respect to the metric tensor gives the retarded two-point function in \textit{Minkowski} space-time \cite{Romatschke:2017ejr} 
\begin{equation}
  \label{eq:gr}
G_R^{xy,xy}(\omega,p\hat{\bf z})=P-i \eta \omega+\omega^2\left(\eta \tau_\pi-\frac{\kappa}{2}+\kappa^*\right)- \frac{p^2 \kappa}{2}+\ldots\,,
\end{equation}
where $P$ is the pressure, $\eta$ is the shear viscosity coefficient, and $\tau_\pi,\kappa,\kappa^*$ are second-order transport coefficients. Note the dual role of $\kappa,\kappa^*$ in curved space-time Eqns. (\ref{eq:t2}) and flat space-time (\ref{eq:gr}) is similar to the Einstein relations for diffusion and conductivity.
Knowledge of $G_R$ at vanishing external frequency $\omega$, but finite wavenumber $p$ is sufficient to determine $\kappa$ \cite{Baier:2007ix,Moore:2012tc,Kovtun:2018dvd}.

I choose to calculate this correlator for the massless O(N) model in 3+1 dimensions. In curved space-time, the action for this theory is given by \cite{Parker:2009uva}
\begin{equation}
  \label{eq:ac}
 \frac{1}{2}\int d^4x\sqrt{-g}\left[g^{\mu\nu}\partial_\mu \vec{\phi}\partial_\nu \vec\phi+\xi R \vec\phi^2-\frac{\lambda}{N}\left(\vec{\phi}^2\right)^2\right]\,,
\end{equation}
where $\vec{\phi}=\left(\phi_1,\phi_2,\ldots\phi_N\right)$ is an N-component scalar field. Here
$\xi$ is a parameter which takes the value $\xi=\frac{1}{6}$ for a conformally coupled scalar.
Calculating $G_R^{xy,xy}$ by varying the energy-momentum tensor for (\ref{eq:ac}) with respect to the metric, the coefficient proportional to $p^2$ in (\ref{eq:gr}) receives two contributions that can be expressed  in terms of \textit{Euclidean} two-point correlation functions \cite{Kovtun:2018dvd},
\begin{equation}
  \label{eq:actualk}
  \kappa=\lim_{p\rightarrow 0}\frac{\partial}{\partial p^2}\left(\langle T^{xy} T^{xy}\rangle_E(p)+ \frac{\xi p^2}{2} \langle \vec{\phi}^2\rangle_E(0) \right)\,.
\end{equation}
Here $\langle\cdot\rangle_E$ denotes Euclidean correlation functions, e.g.
those calculated in a spacetime $S^1\times\mathbb{R}^3$ where
one direction has been compactified on a circle of radius $\beta=T^{-1}$, as in standard thermal quantum field theory \cite{Laine:2016hma}. The corresponding Euclidean Lagrangian is given by $
%
  {\cal L}=\frac{1}{2}\left(\partial_\mu \vec{\phi}\right)\cdot\left(\partial_\mu \vec{\phi}\right)+\frac{\lambda}{N}\left(\vec{\phi}^2\right)^2
  $
and the energy-momentum tensor component by $T^{xy}=\partial^x \vec{\phi} \partial^y \vec{\phi}$.
The Euclidean correlator in (\ref{eq:actualk}) thus becomes\footnote{Usually a whole chain of one-loop diagrams contributes to two-point correlators at large N. However, for the $T^{xy}T^{xy}$ correlator considered here, the presence of the momenta $k^x,k^y$ implies that after momentum-space integration, only the single-loop contribution survives.}
\begin{equation}
  \label{eq:ecorr}
  \langle T^{xy}T^{xy}\rangle_E=2 N \sumint_K k_x^2 k_y^2 \Delta(\omega_n,{\bf k}) \Delta(\omega_n,{\bf k}-{\bf p})\,,
  \end{equation}
where $\Delta(\omega_n,{\bf k})$ is the full two-point function of the scalar field,
\begin{equation}
  \frac{\int {\cal D}\phi e^{-\int d^4x {\cal L}}\phi_i(x) \phi_j(0)}{\int {\cal D} \phi e^{-\int d^4x {\cal L}}}=\delta_{ij}\Delta(x)\,.
\end{equation}

\begin{figure*}[t]
  \includegraphics[width=0.7\linewidth]{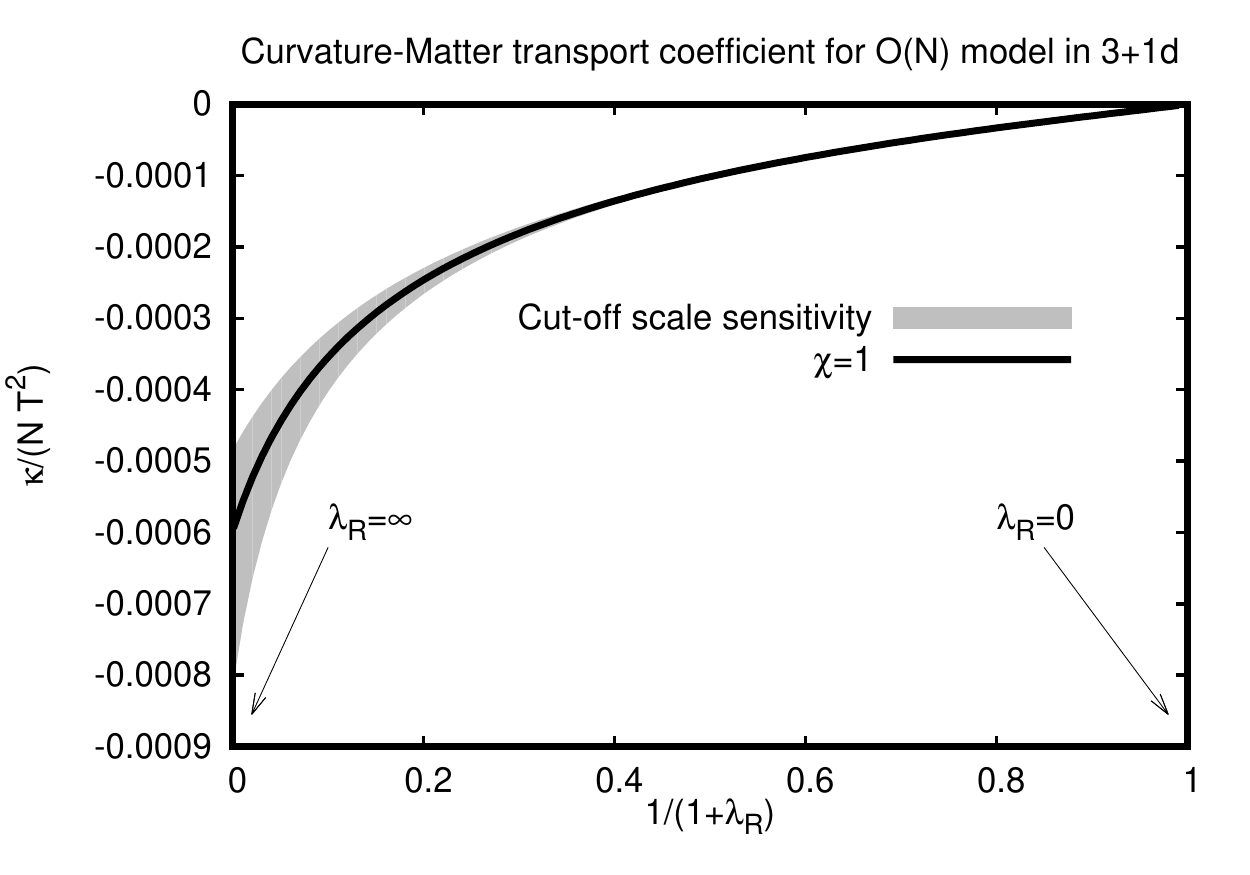}
  \caption{\label{fig1} Transport coefficient $\kappa$  for the O(N) model in 3+1 dimensions in the large N limit as a function of coupling. Results are shown using a compactified interval $\frac{1}{1+\lambda_R}\in [0,1]$ in order to show all coupling values. Arrows indicate free theory  and strong coupling limits, respectively. The band was generated by varying the scale through choosing $\chi \in [\frac{1}{2},2]$ in (\ref{eq:lp}), in turn  quantifying the sensitivity of $\kappa$ to the cut-off scale. See text for details.}
\end{figure*}

Introducing an auxiliary field $\sigma=\frac{\vec{\phi}^2}{N}$ and Lagrange multiplier $\zeta$ and subsequently integrating out $\sigma$, the partition function $Z=\int {\cal D} \phi e^{-\int d^4x {\cal L}}$  can be rewritten as
\begin{equation}
Z=\int {\cal D} \phi {\cal D}\zeta e^{-\int d^4x \left[\frac{1}{2}\vec{\phi}\left(-\Box +i \zeta\right)\vec{\phi}+ \frac{N \zeta^2}{16\lambda} \right]}\,.
\end{equation}
In the large N limit, only the zero mode $\zeta_0$ contributes, and as a consequence the partition function can be calculated exactly from the location of the saddle at $i\zeta_0=z^*$,
\begin{eqnarray}
  \label{eq:parti}
  Z&=&\sqrt{\frac{\beta V N}{16 \lambda \pi}}\int d\zeta_0 e^{-\beta V N \left[\frac{\zeta_0^2}{16\lambda}+J\left(\sqrt{i \zeta_0}\right)\right]}\,,\nonumber\\
  &=&e^{-\beta V N\left[J\left(\sqrt{z^*}\right)-\frac{z^{* 2}}{16\lambda}\right]}\,,
\end{eqnarray}
where $V$ is the volume of ${\mathbb R}^3$, $J(m)\equiv T \sum_{\omega_n}\mu^{2\epsilon}\int\frac{d^{3-2\epsilon}}{(2\pi)^{3-2\epsilon}}\ln\left(\omega_n^2+m^2\right)$ in dimensional regularization \cite{Laine:2016hma} and $\omega_n=2\pi n T$ are the bosonic Matsubara frequencies. (Note that this is completely analogous to the case of 2d and 3d discussed in Refs.~\cite{Romatschke:2019rjk,Romatschke:2019wxc,Romatschke:2019ybu}.)
The two-point function thus becomes
\begin{equation}
  \label{eq:prop}
  \Delta(\omega_n,{\bf k})=\frac{1}{\omega_n^2+{\bf k}^2+z^*}\,,
\end{equation}
where the location of the saddle $z^*$ is given as the solution of the non-perturbative ``gap-equation''
\begin{equation}
  \label{eq:gap}
  z^*=4\lambda I(\sqrt{z^*})\,.
\end{equation}
Here $I(m)=2\frac{d J(m)}{dm^2}=\sumint\, _{k} [\omega_n^2+{\bf k}^2+m^2]^{-1}$ is a standard thermal integral found in textbooks such as Ref.~\cite{Laine:2016hma}
\begin{equation}
  \label{eq:Idef}
I(m)  =-\frac{m^2}{16 \pi^2 \epsilon}-\frac{m^2}{16 \pi^2}\ln\frac{\bar \mu^2 e^1}{m^2}+\frac{m T}{2 \pi^2}\sum_{n=1}^\infty \frac{K_1\left(\frac{n m}{T}\right)}{n}\,,
  \end{equation}
where $\bar\mu^2=4 \pi \mu^2 e^{-\gamma_E}$ is the renormalization scale parameter in the $\overline{\rm MS}$ scheme.
Inspecting (\ref{eq:gap}), one can non-perturbatively renormalize the theory by introducing a renormalized coupling constant $\lambda_R$ as
\begin{equation}
  \label{eq:ren}
  \frac{1}{\lambda_R}=\frac{1}{\lambda}+\frac{1}{4\pi^2 \epsilon}\,.
\end{equation}

This renormalization condition implies a positive $\beta$-function for all couplings. Integrating up the renormalization group equation gives
\begin{equation}
  \frac{1}{\lambda_R(\bar\mu)}=\frac{1}{4\pi^2}\ln \frac{\Lambda_{LP}^2}{\bar\mu^2}\,,
  \end{equation}
where $\Lambda_{LP}$ is the Landau pole of the theory (defined as the scale where $\lambda_R(\Lambda_{LP})=\infty$). 

Expressing the thermal mass in (\ref{eq:prop}) as $z^*=m_B^2 T^2$, the dimensionless parameter $m_B$ is determined from (\ref{eq:gap}) as
\begin{equation}
  \label{eq:gapr}
m_B=\frac{8 \sum_n \frac{K_1\left(n m_B\right)}{n}}{\ln\frac{\Lambda_{LP}^2 e^1}{m_B^2 T^2}}=\frac{8 \sum_n \frac{K_1\left(n m_B\right)}{n}}{\frac{4\pi^2}{\lambda_R}+\ln\frac{\bar\mu^2 e^1}{m_B^2 T^2}}\,,
\end{equation}
either in terms of the ratio $\Lambda_{LP}/T$ or in terms of the renormalized running coupling. (Note that $m_B$ is independent from the choice of the renormalization scale parameter $\bar\mu$ as it should be for a physical observable.)

Note that while the gap equation (\ref{eq:gapr}) formally is well-defined for all temperature scales $T\in [0,\frac{\Lambda_{LP}}{2}]$, close to the Landau pole there will be modifications arising from radiative corrections to the effective theory Lagrangian. (This may be verified explicitly by adding a term such as $\frac{\lambda_2}{\Lambda_{LP}^2}\left(\vec{\phi}^2\right)^3$ to the Lagrangian (\ref{eq:ac}), which is allowed for an UV-incomplete theory.) It is possible to test for the sensitivity to the cut-off scale by e.g. choosing units as
\begin{equation}
\label{eq:lp}
2 \pi T=\frac{\Lambda_{LP} e^{-\frac{2\pi^2}{\lambda_R}}}{\chi}\,,
  \end{equation}
with fixed $\lambda_R$ and varying $\chi\in[\frac{1}{2},2]$.

In practice, the renormalized gap equation (\ref{eq:gapr}) possesses two solutions for $m_B$. Only the smaller one of these corresponds to a local minimum of the exponent, thus the larger one will be discarded in the following. The solution $m_B$ then fixes the form of the two-point function (\ref{eq:prop}) non-perturbatively, and in turn allows calculation of the transport coefficient $\kappa$ from (\ref{eq:actualk}). Specifically, performing the angular averages in (\ref{eq:ecorr}) leads to
\begin{equation}
\label{eq:kd2}
\frac{\kappa}{2N}\equiv \frac{2}{105} \sumint_K \left(4 k^6 \Delta^4(\omega_n,k)-7 k^4 \Delta^3(\omega_n,k)\right)+\frac{\xi}{2} I(m_B T)\,.
\end{equation}
Inspecting this equation, one notices that the last term is divergent for $\epsilon\rightarrow 0$, cf.~(\ref{eq:Idef}). Therefore, unless this divergence is exactly canceled by the other contributions, the result for $\kappa$ is meaningless. Using $\frac{\partial}{\partial m^2}\Delta=-\Delta^2$ repeatedly and performing a standard thermal sum, one finds
\begin{eqnarray}
 A= \sumint_K k^6 \Delta^4(\omega_n,k)&=&-\frac{\partial^3}{(\partial z^*)^3}\int_k \frac{k^6}{12 E_k}\left(1+2 n_B(E_k)\right)\,,\nonumber\\
 B= \sumint_K k^4 \Delta^3(\omega_n,k)&=&\frac{\partial^2}{(\partial z^*)^2}\int_k \frac{k^4}{4 E_k}\left(1+2 n_B(E_k)\right),\,\,\,
\end{eqnarray}
where $E_k=\sqrt{k^2+z^*}$ and $n_B(x)=\frac{1}{e^{x/T}-1}$. Expanding $n_B(x)=\sum_{n=1}^\infty e^{-n \beta x}$ both of the above integrals can be evaluated analytically, finding
\begin{eqnarray}
4A&=&-\frac{35 z^*}{64 \pi^2}\left(\frac{1}{\epsilon}+\frac{37}{105}+\ln \frac{z^*}{\bar\mu^2}-8\sum_{n=1}^\infty \frac{K_1(n \beta \sqrt{z^*})}{n \beta}\right)\nonumber\,,\\
7 B&=&-\frac{105 z^*}{128 \pi^2}\left(\frac{1}{\epsilon}+\frac{1}{15}+\ln\frac{z^*}{\bar\mu^2}-8\sum_{n=1}^\infty \frac{K_1(n \beta \sqrt{z^*})}{n \beta}\right)\,.\nonumber
\end{eqnarray}
Inserting these results into (\ref{eq:kd2}), I find that the $\epsilon\rightarrow 0$ divergence as well as the sums over Bessel functions both cancel for $\xi=\frac{1}{6}$, giving rise to the finite and simple result
\begin{equation}
\label{eq:fink}
\kappa=-\frac{13 N T^2 m_B^2}{2520 \pi^2}\,,
\end{equation}
with $m_B$ given by the solution of (\ref{eq:gapr}).
This is the main result of this work. A quick cross-check reveals that in the free theory limit $\lambda_R\rightarrow 0$ Eq.~(\ref{eq:gapr}) gives $m_B\rightarrow 0$, so that $\lim_{\lambda_R\rightarrow 0}\kappa=0$ (matching the result found in Ref.~\cite{Kovtun:2018dvd} for a conformally coupled scalar).

\begin{figure}[t]
  \includegraphics[width=0.95\linewidth]{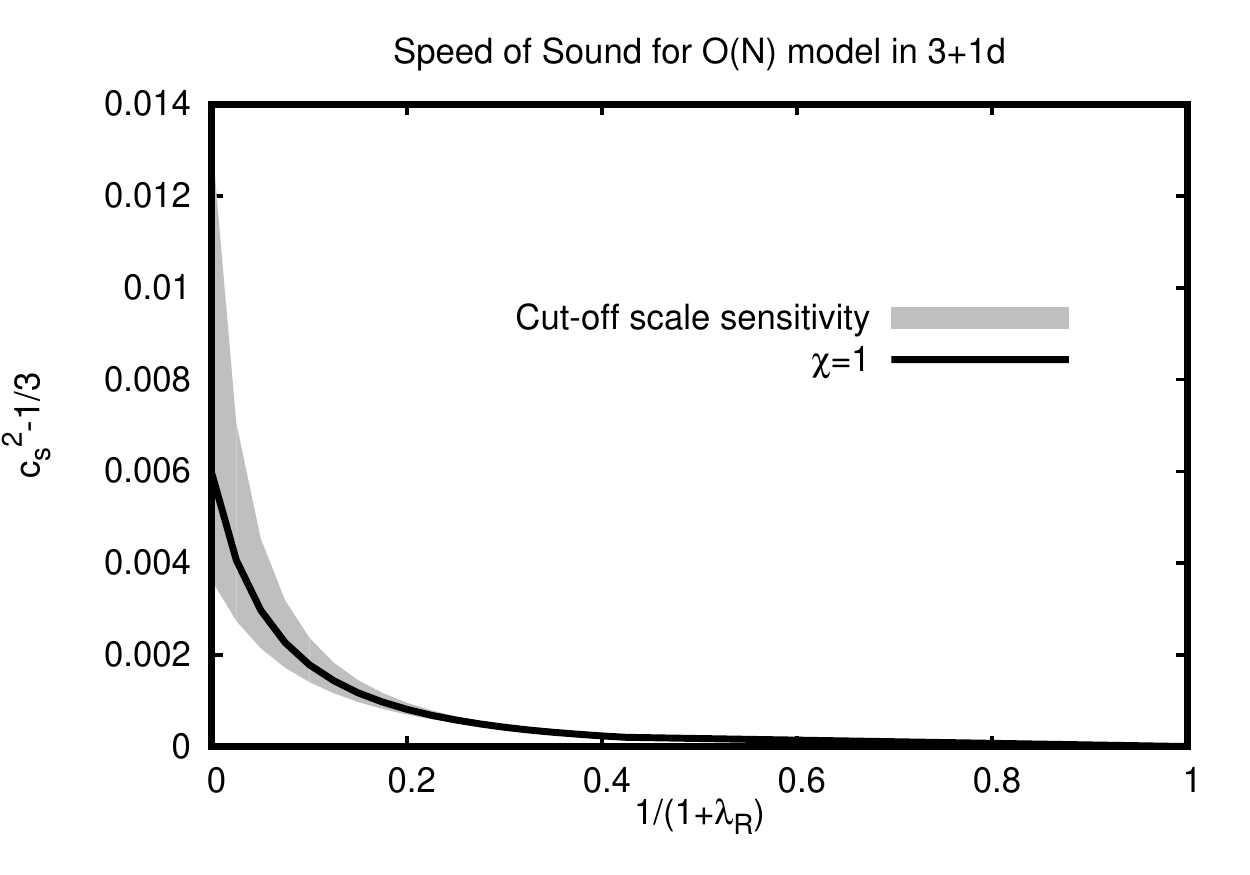}
  \caption{\label{fig2} Speed of sound squared for the O(N) model in 3+1 dimensions in the large N limit as a function of coupling. Results are shown using a compactified interval $\frac{1}{1+\lambda_R}\in [0,1]$ in order to show all coupling values.  The band was generated by varying the scale  through choosing $\chi \in [\frac{1}{2},2]$ in (\ref{eq:lp}), in turn  quantifying the sensitivity of $c_s^2$ to the cut-off scale. }
\end{figure}

Of course, also thermodynamic properties of the O(N) model in 3+1 dimensions may be evaluated non-perturbatively along the same lines. For instance, the pressure (minus the free energy) is found from (\ref{eq:parti}) as $P=\frac{\ln Z}{\beta V}$. It is worth pointing out that -- using the explicit result \cite{Laine:2016hma} for $J(m)=-\frac{m^4}{64\pi^2}\left[\frac{1}{\epsilon}+\ln \frac{\bar\mu^2 e^{3/2}}{m^2}\right]-\frac{m^2 T^2}{2\pi^2}\sum_n \frac{K_2\left(n \beta m\right)}{n^2}$ -- the non-perturbative coupling renormalization (\ref{eq:ren}) is sufficient to remove all divergences in the pressure (cf. Ref.~\cite{Blaizot:2000fc}), so that no counterterm for the cosmological constant is required. This leads to 
\begin{equation}
  P=\frac{N T^4}{64\pi^2}\left[m_B^4 \ln \frac{\Lambda_{LP}^2 e^{3/2}}{m_B^2 T^2}+32 m_B^2 \sum_{n=1}^\infty \frac{K_2(n m_B)}{n^2}\right]\,.
  \end{equation}
The entropy density $s\equiv \frac{\partial P}{\partial T}$ may be obtained most easily from (\ref{eq:parti}) and (\ref{eq:gap}), so that contributions proportional to $\frac{\partial z^*}{\partial T}$ cancel, leading to the result
\begin{equation}
  \label{eq:entropy}
  s=\frac{N T^3 m_B^3}{2\pi^2} \sum_{n=1}^\infty \frac{K_3(n m_B)}{n}\,.
\end{equation}
For weak coupling where $m_B\rightarrow 0$, $s\rightarrow s_{\rm free}=\frac{4 N T^3 \pi^2}{90}$, the well-known Stefan-Boltzmann result for a free theory. 

From the thermodynamic relation $\epsilon+P=s T$ and this result, the trace anomaly can be evaluated to be
\begin{equation}
 \epsilon-3P=-\frac{N T^4 m_B^4}{32 \pi^2}\,.
\end{equation}
Note that the result is negative and that most contributions have canceled because of the gap equation (\ref{eq:gapr}). Finally, the speed of sound squared can be calculated as $c_s^2=\frac{s/T}{\frac{\partial s}{\partial T}}$, and evaluated numerically, see Fig.~\ref{fig2}. Note that the speed of sound is very close to (and above) the conformal result $c_s^2=\frac{1}{3}$, which indicates that the O(N) model, though not a conformal theory (CFT), is numerically very close to a CFT for most coupling values. Indeed, it has not escaped my attention that the ratio $s/s_{\rm free}$ calculated from (\ref{eq:entropy}) seems to go to a constant value of approximately 85 percent for $\lambda_R\rightarrow \infty$ and $\chi=1$, very much in line with the universal strong-weak thermodynamic behavior found in 2+1d CFTs \cite{Romatschke:2019ybu,DeWolfe:2019etx}.

One referee of this work remarked that for the O(N) model, $c_s^2\geq \frac{1}{3}$ while the authors of Ref.~\cite{Hohler:2009tv} found that $c_s^2\leq \frac{1}{3}$ for a class of holographic theories. The apparent discrepancy can be understood to originate from the different sign of the $\beta$-function in scalar field theories (such as the O(N) model) and non-abelian gauge theories such as those considered in Ref.~\cite{Hohler:2009tv}. To take a simple example, let us consider the relation between the pressure and energy density in weakly coupled single-component $\phi^4$ theory, which can be obtained by combining results found in Ref.~\cite{Laine:2016hma} to give
\begin{equation}
  \label{eq:sos}
  P=\frac{\epsilon}{3}+\frac{T^4}{576} \frac{\partial \lambda}{\partial \ln T}+\ldots\,,
\end{equation}
where $\frac{\partial \lambda}{\partial \ln T}$ is the $\beta$ function of $\phi^4$ theory. Since in scalar field theories, $\frac{\partial \lambda}{\partial \ln T}>0$, Eq.~(\ref{eq:sos}) implies $c_s^2\equiv \frac{dP}{d\epsilon}\geq \frac{1}{3}$, while for asymptotically free theories with negative $\beta$ function $c_s^2\leq \frac{1}{3}$.

\section{Discussion and Conclusions}

The transport coefficient $\kappa$ given in (\ref{eq:fink}) may be evaluated for any value of the renormalized coupling $\lambda_R$ by solving (\ref{eq:gapr}) numerically. The sensitivity to cut-off scale effects may be tested by the choice (\ref{eq:lp}) through varying $\chi$. Results for $\kappa$ for all couplings are shown in Fig.~\ref{fig1}. From this figure, it can be seen that cut-off scale sensitivity is minor (smaller than 10 percent) for $\lambda_R\lesssim 2.45$ and less than a factor of two even for $\lambda_R\rightarrow \infty$. This compares favorably with the situation found for the QCD shear viscosity calculated to NLO in perturbation theory \cite{Ghiglieri:2018dib}. The weak sensitivity to cut-off scale effects suggests that the result (\ref{eq:fink}) constitutes an example of a transport coefficient that is known non-perturbatively for a large range of coupling values. As such, this example may be useful for instance for testing approximation techniques (either at weak or at strong coupling), or conceivably in early-time cosmology where curvature matter couplings will play an important role in the dynamics.

The results found in this work are exact only in the strict large N limit. However, based on the non-perturbative results for scalar theory in 1+1d in Ref.~\cite{Romatschke:2019wxc}, I conjecture that the large N results in this work constitute quantitatively good approximations for finite N at arbitrary coupling, including N=1 scalar $\phi^4$ theory.

Is it possible to non-perturbatively evaluate other transport coefficients in a similar manner? The answer to this question likely is affirmative since other channels of the energy-momentum tensor two-point function couple to transport coefficients such as $\kappa^*,\xi_5,\xi_6$ in a similar manner, cf. Refs.~\cite{Romatschke:2017ejr,Moore:2012tc,Kovtun:2018dvd} for details. 

It would also be interesting to extend this work on the O(N) model and study non-perturbative relations between $\kappa,\kappa^*$ and the shear viscosity coefficient discussed in Ref.~\cite{Kleinert:2016nav}, or the Haack-Yarom relation \cite{Haack:2008xx}, which has been investigated extensively in holographic theories \cite{Grozdanov:2014kva}.

It is my hope that this work could instill interest and further progress in the field of non-perturbative transport calculations. 

  \section{Acknowledgments}

  This work was supported by the Department of Energy, DOE award No DE-SC0017905. I would like to thank S.P. de Alwis, T. DeGrand, S. Grozdanov, P. Kovtun and G. Moore for helpful discussions.

\bibliography{kappa}

\end{document}